\documentclass[aps,pre,twocolumn,groupedaddress,showpacs]{revtex4}
\usepackage{graphicx}
\usepackage{amssymb,amsfonts,amsmath}

\newcommand{\ii}{i}          
\newcommand{\cc}{\text{c.c.}}
\newcommand{\hot}{\text{h.o.t.}}
\newcommand{\eps}{\varepsilon}
\newcommand{\bigO}{\ensuremath{{\mathcal{O}}}}

\newcommand{\trave}[1]{\langle (\overline{{#1}^2})^{1/2} \rangle}
\newcommand{\tave}[1]{\langle #1 \rangle}
\newcommand{\rsqave}[1]{(\overline{{#1}^2})^{1/2}}
\newcommand{\rave}[1]{\overline{#1}}
\newcommand{\rms}{\text{rms}}

\begin{document}

\title{Anomalous Scaling on a Spatiotemporally Chaotic Attractor}

 \author{Ralf W. Wittenberg}
 \email{ralf@sfu.ca}
 \author{Ka-Fai Poon}
 \affiliation{Department of Mathematics, Simon Fraser University,
   Burnaby, BC V5A 1S6, Canada}

\date{\today}

  \begin{abstract} 
  The Nikolaevskiy model for pattern formation with continuous symmetry exhibits spatiotemporal chaos with strong scale separation. Extensive numerical investigations of the chaotic attractor reveal unexpected scaling behavior of the long-wave modes.  Surprisingly, the computed amplitude
and correlation time scalings are found to differ from the values obtained by asymptotically consistent multiple-scale analysis.  However, when higher-order corrections are added to the leading-order theory of Matthews and Cox, the anomalous scaling is recovered. 
  \end{abstract}


\maketitle

\section{Introduction}
\label{sec:intro}

The study of dynamics and pattern formation in nonlinear spatially extended dynamical systems continues to attract considerable attention \cite{CrHo93,REW00k,Hoyl06k}, with potential solution behaviors ranging from attracting steady states to complex spatiotemporal dynamics including spatiotemporal chaos (STC), which is characterized by temporally chaotic dynamics with decaying spatial correlations and a finite density of positive Lyapunov exponents \cite{CrHo93,FiEg06}.

In the search for general principles, much effort has gone into the derivation and study of simplified model equations to highlight and clarify pattern-forming properties of more complicated ``full'' systems.  Such paradigmatic models, such as the Ginzburg-Landau equation for the slow evolution of modulations of an underlying pattern, have been remarkably fruitful in predicting the leading-order scaling behavior and stability of patterns, by comparison with numerical or experimental results.

The identification of relevant scaling laws forms an integral part of the asymptotic procedures used to derive such reduced model systems.  The successful application of such methods rests on the expectation that the leading-order truncation obtained subject to a consistent scaling assumption indeed captures the dominant scaling of the full problem, in the sense that the perturbative effects of neglected terms are ``small'' and occur at the scaling implied by the dominant balance.  In this paper, we present an example which appears to challenge this expectation.

It is known from various examples that the stability of finite-wavelength pattern or wave phenomena may be complicated through interaction with a mean flow; for instance, in convection with free-slip boundary conditions \cite{Bern94}, in some reaction-diffusion systems \cite{BKMV95} and elsewhere, such coupling can lead to unusual instability mechanisms of the underlying pattern or flow, whose analysis may display mixing of scales in the perturbation parameter and require higher-order terms in the perturbative expansions for a complete understanding \cite{Trib97}.

In the present work we study a model for one-dimensional pattern formation in the presence of continuous symmetry, for which the interaction of modes with two well-separated length scales leads to spatiotemporal chaos.  Focussing in this case on characterizing the long-term chaotic regime rather than on its onset via the initial stability problem, for the statistics of the long-wave mode we find a scaling discrepancy between predictions of leading-order multiple-scale analysis and computations of the full system.  Unexpectedly, though, the agreement in scaling is restored by including next-order corrections in the amplitude equations.

We investigate the Nikolaevskiy partial differential equation (PDE), given in canonical (derivative) form as 
\begin{equation}
  \label{eq:nik}
  u_t + u u_x = - \partial_x^2 \left[ r - \left( 1 + \partial_x^2 \right)^2 \right] u 
\end{equation}
(where $u_x \equiv \partial_x u \equiv \partial u/\partial x$), which was originally proposed as a model for longitudinal seismic waves in viscoelastic media \cite{Niko89,BeNi93}, and has more recently been obtained in the context of phase dynamics in reaction-diffusion systems \cite{FuYa01,Tana04} and finite-wavelength transverse instabilities of traveling fronts \cite{CoMa07}.  In fact, the Nikolaevskiy equation has a continuous symmetry---in the form \eqref{eq:nik} it is Galilean-invariant, and preserves the spatial mean---and appears to be a paradigmatic model for short-wave pattern formation with such symmetry \cite{Trib97,MaCo00,CoMa07}.

Considering individual Fourier modes of the form $\exp(\sigma t + \ii k x)$, the linear dispersion relation about the trivial solution $u \equiv 0$ is 
$\sigma(k) = k^2 [ r - ( 1 - k^2)^2 ]$.  
Since for $r \leq 0$, all initial conditions of \eqref{eq:nik} decay to a spatially homogeneous solution, in the following we consider only $r > 0$ and write $r = \eps^2$; as usual the $\bigO(\eps)$ width of the unstable band and $\bigO(\eps^2)$ maximum growth rate suggest natural slow length and time scales $X = \eps x$, $T = \eps^2 t$.  
For $0 < r = \eps^2 \ll 1$, the Nikolaevskiy equation \eqref{eq:nik} has  a one-parameter (up to translation) family of stationary periodic solutions, or rolls, $u_q(x)$ for $|q| < 1/2$, which may be found by weakly nonlinear analysis to be \cite{TrVe96,MaCo00}
\begin{equation}
  \label{eq:nikroll}
  u_q(x) = 6 \eps \sqrt{1- 4 q^2} e^{\ii (1 + \eps q) x} + \cc + \bigO(\eps^2) .
\end{equation}

Previous studies of \eqref{eq:nik} have revealed some surprises.  Due to the presence of a symmetry-induced long-wave (Goldstone) mode $U_0$ interacting with the weakly unstable pattern mode $U_1$, all roll equilibria $u_q(x)$ are unstable for all $r>0$ \cite{TrVe96,MaCo00,CoMa07,Trib08}; instead, there is a direct supercritical transition from spatial uniformity to spatiotemporal chaos \cite{TrTs96}.  Similar behavior occurs, for instance, in electroconvection in liquid crystals \cite{RHKP96,HHHKT97}, and has been termed ``soft-mode turbulence'' \cite{Trib97,Trib08}.

By deriving leading-order modulation equations using multiple-scale analysis, Matthews and Cox \cite{MaCo00} furthermore showed that in the chaotic state, $U_1$ (and hence $u$) has a rather unusual $r^{3/4} = \eps^{3/2}$ scaling.  In the present work, through extensive numerical studies we show that Nikolaevskiy STC exhibits yet more unexpected scaling: amplitudes and correlation times of the large-scale mode $U_0$ have an anomalous dependence on $r$, seemingly inconsistent with leading-order multiple-scale analysis, though recovered with the addition of next-order terms to the amplitude equations.  In particular, in the chaotic regime the mean-square amplitude of $U_0$ appears to scale as $r^{7/8}$, rather than the predicted $r^1$.

\section{Numerical methods, solution properties and modal decomposition}
\label{sec:solns}

We numerically computed (mean-zero) solutions of the Nikolaevskiy equation \eqref{eq:nik} on an $\ell$-periodic domain in the chaotic regime, over a wider $\eps$-range than in earlier studies: $10^{-5} \leq \eps^2 \leq 0.2$.  The system size $\ell$ was chosen so that the number of Fourier modes in the unstable band of the dispersion relation $\sigma(k)$, about $\lfloor \eps \ell/2\pi \rfloor$, sufficed to capture the large-system limit; to minimize system size effects across our computations, we fixed the domain length in the slow space variable $X = \eps x$, at $L = \eps \ell = 2\pi \cdot 128 m/10$ with $m$=1 and 2 (about 12 and 25 unstable modes, respectively), and checked our results with $m$=4.  We used a pseudo-spectral method in space with $N = 2^J$ Fourier modes for $J$=11 to 16, chosen so that the maximum wave number retained was well within the strongly decaying regime of $\sigma(k)$: $k_m = N \cdot 2\pi/\ell \geq 6$.  Our time integration was performed using a fourth-order Runge-Kutta exponential time differencing (ETDRK4) scheme \cite{CoMa02,KaTr03}, with a typical time step of $\Delta t = 0.02/\eps^2$, fixed in the slow time variable $T = \eps^2 t$; after a transient, we computed statistics over a time interval of length $t_m = T_m/\eps^2 = 4\cdot 10^4 /\eps^2$.

The (time-averaged) power spectrum $S(k) = \tave{|\hat{u}_k|^2}$ of Fig.~\ref{fig:nikps} captures the dominant features of the numerically observed chaotic dynamics:%
\begin{figure}[tb]
  \begin{center}
        \includegraphics[width = 3.1in]{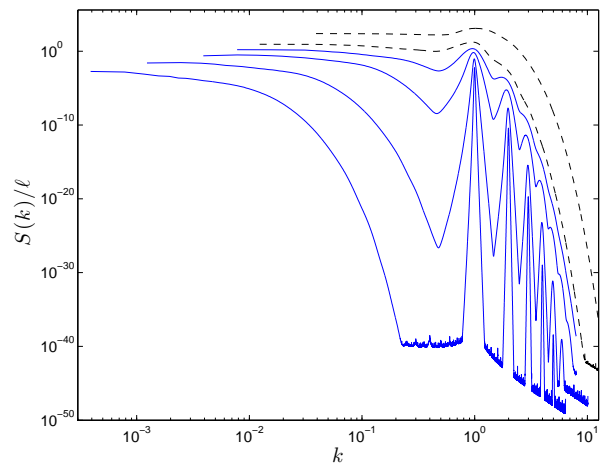}
    \caption{(Color online) Time-averaged Fourier power spectrum $S(k)/\ell$ for the Nikolaevskiy equation \eqref{eq:nik} for $L = \eps \ell = 2\pi \cdot 25.6$ and various $\eps^2$ values.  From top to bottom, the spectra shown are for $\eps^2 = 1.0$, $0.1$, $0.04$, $0.01$, $10^{-3}$, and $10^{-4}$.  Note the separation of scales for $\eps^2 \lesssim 0.04$ (blue solid curves).}
    \label{fig:nikps}
  \end{center}
\end{figure}
As pointed out by Tanaka \cite{Tana05}, for ``large'' $\eps$ the STC appears qualitatively similar to that of the related well-known Kuramoto-Sivashinsky equation \cite{WiHo99}; Fourier modes with $0<k<1$ are (at most) weakly damped, and the power spectrum reveals no significant scale separation for $\eps^2 \gtrsim 0.1$ (dashed curves).

However, for sufficiently ``small'' $\eps$, we observe the distinct gap between the $k \approx 0$ and $|k| \approx 1$ energetic modes which characterizes ``Nikolaevskiy chaos'' \cite{Tana05}.  Note that for $\eps \ll 1$, the subdominant peaks at $|k| \approx 2, 3, \dots$ represent higher harmonics slaved to the Fourier modes $|k| \approx 1$ via the quadratic nonlinearity; we remark that exponential time differencing integration schemes \cite{CoMa02} are particularly well-suited to capturing this slaving behavior and improving the accuracy at small scales.

\begin{figure}[tb]
  \begin{center}
    \includegraphics[width = 3.2in]{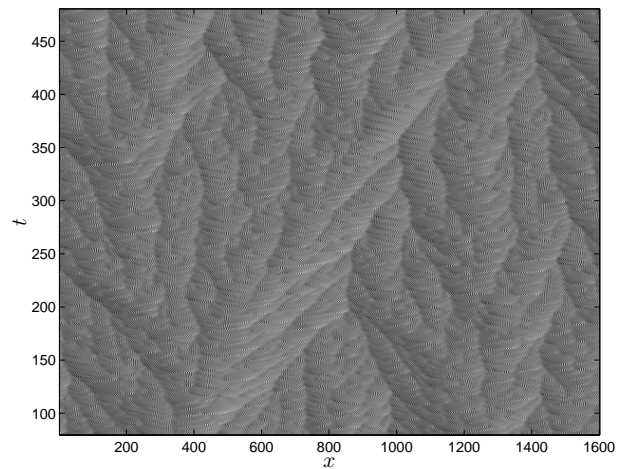}
    \caption{Gray-scale visualization of a computed solution of the Nikolaevskiy equation \eqref{eq:nik} for $r = \eps^2 = 0.04$, with domain size $\ell = 1600$ and over a time interval of length $t = 400$; since the roll wavelength $\lambda \approx 2\pi \ll \ell$, the small-scale rolls are barely visible in this representation.}
    \label{fig:nikxt}
  \end{center}
\end{figure}

\begin{figure}[tb]
  \begin{center}
  \includegraphics[width=3.2in]{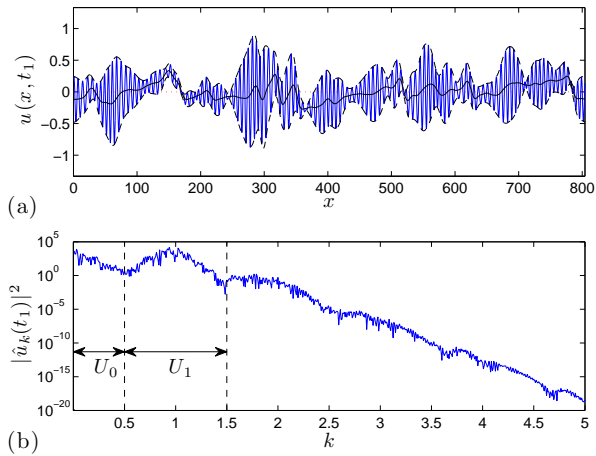} 
    \caption{(Color online) (a) Snapshot at a fixed time $t_1$ of a solution $u(x,t_1)$ (blue) of the Nikolaevskiy equation
      \eqref{eq:nik} with $r = \eps^2 = 0.04$ and domain length $\ell =
      800$.  The figure also shows the
      reconstructed mean mode $U_0$ (black) and envelope $U_0 \pm 2 |U_1|$ (black dashed line).
      (b) Instantaneous spectrum $|\hat{u}_k(t_1)|^2$ of the
      solution in (a), showing the wavenumber ranges for the Fourier
      filters.}
    \label{fig:nikux}
  \end{center}
\end{figure}

The strong scale separation of the spatiotemporally chaotic dynamics in this small-$\eps$ regime is readily apparent in a representation of a typical numerically computed time series on the multiple-scale chaotic attractor of the Nikolaevskiy PDE \eqref{eq:nik}, as in Fig.~\ref{fig:nikxt},
or of a snapshot of a typical solution as seen in Fig.~\ref{fig:nikux}(a):
Solutions on the attractor are characterized by long-wave modulations, aperiodic in space and time, of small-scale rolls of wavelength $\lambda = 2\pi (1 + \bigO(\eps))$ (corresponding to the linearly unstable modes of $\sigma(k)$).
This motivates a natural decomposition of the solution $u(x,t)$ of the Nikolaevskiy PDE \eqref{eq:nik}: with $X = \eps x$, $T = \eps^2 t$, we have
\begin{equation}
  \label{eq:moddesc}
  u(x,t) = U_1(X,T) e^{\ii x} + \cc + U_0(X,T) + \hot ,
\end{equation}
where $U_1$ and $U_0$ are slowly varying envelopes of the dynamically significant Fourier modes near $|k|=1$ and $k=0$, respectively.

We have implemented Fourier filters to extract $U_0$ and $U_1$ at each time step from the Fourier space numerical solution $\hat{u}_k(t)$ of the full PDE \eqref{eq:nik}; see Fig.~\ref{fig:nikux}(b): For some $k_* \in (0,1)$, we identify the modes with $|k| < k_*$ as being ``large-scale'', while those with $|k| > k_*$ belong to the patterned mode and its higher harmonics.  In view of the strong suppression of the linearly damped modes near $k = 0.5$ for $\eps^2 \lesssim 0.04$ (see Fig.~\ref{fig:nikps}), we may safely choose $k_* = 0.5$; we have verified that our results are insensitive to the choice of $k_*$.  Thus we obtain $U_0$ from $\{ \hat{u}_k | \ |k| < 0.5\}$, and $U_1 e^{\ii x}$ and hence $U_1$ from $\{ \hat{u}_k | \ 0.5 < k < 1.5\}$.  In Fig.~\ref{fig:nikux}(a) we have also shown the mean mode $U_0$ and the reconstructed envelope $U_0 \pm 2 |U_1|$.

\section{Modulation equations}
\label{sec:modulation}

In the light of the decomposition \eqref{eq:moddesc} for solutions of \eqref{eq:nik}, one may seek a modulation theory for the slow evolution of the envelopes, coupling the amplitude $U_1$ of the finite-wavelength spatial pattern mode with the long-wave mode $U_0$ induced by the continuous symmetry \cite{CoFa85,Malo92,MaCo00}.  
In this context it is of particular interest to determine the exponents $\alpha$ and $\beta$ describing the scaling of $U_1$ and $U_0$ as $\eps \to 0$.
Thus one postulates the Ansatz
\begin{equation}
  \label{eq:genscal}
  u(x,t) \sim \eps^{\alpha} A(X,T) e^{\ii x} + \cc + \eps^{\beta} f(X,T) + \dots,
\end{equation}
where $A$ and $f$ are assumed $\bigO(1)$, and $f$ has mean zero.

The $\bigO(\eps)$ amplitude of the known roll solutions $u_q(x)$ \eqref{eq:nikroll} suggests the apparently natural choice $\alpha$=1 (see for instance \cite{Malo92}), familiar from other pattern-forming systems \cite{CrHo93}, which we will denote the Ginzburg-Landau (GL) scaling.  It turns out, though, as discussed in \cite{MaCo00}, that amplitude equations for \eqref{eq:nik} derived using $\alpha$=1 in \eqref{eq:genscal} are asymptotically inconsistent for any choice of $\beta$ if $f \not= 0$.

Instead, as shown by Matthews and Cox \cite{MaCo00}, the only asymptotically consistent amplitude equations are obtained by using the Ansatz \eqref{eq:genscal} with $\alpha$=3/2, $\beta$=2; we shall call this the MC scaling.  Standard multiple-scale techniques then yield the leading-order amplitude equations proposed to describe STC in the Nikolaevskiy equation \eqref{eq:nik}: the Matthews-Cox (MC) equations \cite{MaCo00} are 
\begin{eqnarray}
  \label{eq:MC1}
  A_T & = & A + 4 A_{XX} - \ii f A , \\
  \label{eq:MC2}
  f_T & = & f_{XX} - |A|^2_X .
\end{eqnarray}
An immediate prediction of \eqref{eq:genscal} with the MC exponents, that the amplitude of solutions $u(x,t)$ of \eqref{eq:nik} (computed as the time-averaged root-mean-square (rms) amplitude $\trave{u}$, where $\tave{\cdot}$ and $\rave{\ \cdot\ }$ denote time and space averages, respectively) should scale as $\trave{u} = \bigO(\eps^{3/2}) = \bigO(r^{3/4})$, was verified in \cite{MaCo00,Tana05}; see \cite{FHY03} for an analogous result in two space dimensions.

We have performed the multiple-scale analysis of \eqref{eq:nik} to the next order in $\eps$, using the Ansatz \eqref{eq:genscal} with the MC scaling $\alpha$=3/2, $\beta$=2, to obtain the asymptotically consistent $\bigO(\eps)$ amplitude equations
\begin{eqnarray}
   A_T & = & A + 4 A_{XX} - i f A - \eps \left[ \frac{|A|^2 A}{36} +
     \left( f A \right)_X \right] \nonumber \\
   & & - 2 i \eps \left[ A_X + 6 A_{XXX} \right], \label{eq:MCe1} \\
   f_T & = & f_{XX} - |A|^2_X - \eps f f_X . \label{eq:MCe2}  
\end{eqnarray}
(the linear $\bigO(\eps)$ terms in \eqref{eq:MCe1} correspond to next-order corrections to the exact dispersion relation $\sigma(k)$).
Observe that unlike the $\bigO(1)$ equations \eqref{eq:MC1}--\eqref{eq:MC2}, these $\bigO(\eps)$ equations also capture the roll solutions $u_q(x)$ \eqref{eq:nikroll} of the GL regime, which have size $|A| = \bigO(\eps^{-1/2})$ in this scaling.  
One would expect, though, that in the chaotic MC regime (in which $u = \bigO(\eps^{3/2})$, $|A| = \bigO(1)$), the $\bigO(\eps)$ correction terms in \eqref{eq:MCe1}--\eqref{eq:MCe2} would lead to small $\bigO(\eps)$ perturbations to the solutions and their scaling.

\section{Numerical computation of scaling exponents}
\label{sec:exponents}

Following previous studies \cite{MaCo00,Tana05}, we set out to verify that $U_1$ and $U_0$ indeed scale for small $\eps$ as predicted by the MC exponents $\alpha$=3/2, $\beta$=2.  
With a view to isolating and thereby computing the individual scaling exponents directly, Tanaka \cite{Tana05} numerically integrated the PDE \eqref{eq:nik} on a domain of length $\ell$=512 for $\eps^2 \in [0.001,0.4]$, and computed the $\rms$ scaling of the Fourier coefficients nearest in his system to $k$=0 and $k$=1; these Fourier modes were used as proxies for $U_0$ and $U_1$ to argue that the results were consistent with MC scaling.  However, the numerical evidence of \cite[Fig.~5]{Tana05} does not appear sufficiently well-averaged to establish the asserted $\bigO(\eps^2)$ large-scale scaling convincingly.

\begin{figure}[bt]
  \begin{center}
  \includegraphics[width=3.2in]{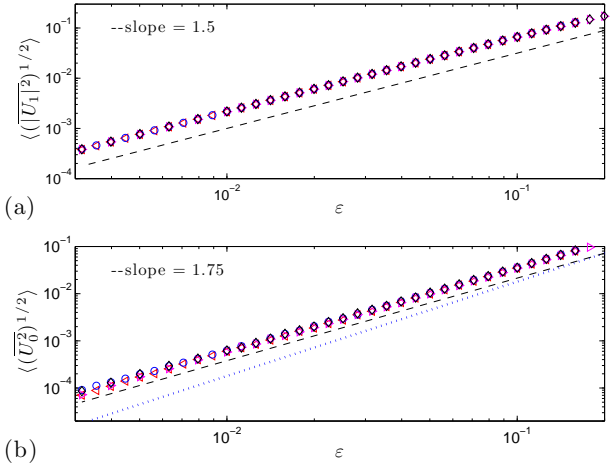} 
    \caption{(Color online) Matthews-Cox scaling of the pattern mode $U_1$, and anomalous scaling of the large-scale mode $U_0$:\\
    (a) Scaling of averaged rms amplitudes of $|U_1|$ from \eqref{eq:nik} and $|\eps^{3/2} A|$ from \eqref{eq:MCe1}--\eqref{eq:MCe2}; we find $\trave{|U_1|} \sim \eps^{3/2} \trave{|A|} \sim \eps^{\alpha'}$ for $\alpha' = 1.49 \pm 0.01$.  The (black) dashed line indicates the MC scaling prediction $\alpha = 3/2$. \\
    (b)  Scaling of averaged rms amplitudes of $U_0$ from \eqref{eq:nik} and $\eps^2 f$ from \eqref{eq:MCe1}--\eqref{eq:MCe2}; we find $\trave{U_0} \sim \eps^2 \trave{f} \sim \eps^{\beta'}$ for $\beta' = 1.75 \pm 0.03$ (black dashed reference line).  
    The (blue) dotted line indicates the MC prediction $\beta = 2$.\\
          Symbols for this and the next figures (hardly distinguishable here): (Magenta and red) Triangles: $L = \eps \ell = 2\pi \cdot 12.8$; (blue) Circles and (black) Diamonds: $L = \eps \ell = 2\pi \cdot 25.6$.  $\triangleright$, $\diamond$: Nikolaevskiy PDE \eqref{eq:nik}, $\ell = 2\pi m \cdot 12.8/\eps$, $m$=1 and 2; $\triangleleft$, $\circ$: $\bigO(\eps)$ amplitude equations \eqref{eq:MCe1}--\eqref{eq:MCe2}, $L = 2\pi m \cdot 12.8$, $m$=1 and 2.}
    \label{fig:U1U0rms}
  \end{center}
\end{figure}

Our implementation of Fourier filters to extract the full slowly-varying envelopes $U_0$ and $U_1$ from the computed solution $u(x,t)$ of the Nikolaevskiy PDE \eqref{eq:nik} provides a more powerful method to estimate the critical exponents, via the $\eps$-dependence of the time-averaged $\rms$ magnitudes of $U_0$, $U_1$.  In Fig.~\ref{fig:U1U0rms} we plot $\trave{|U_1|}$ and $\trave{U_0}$ computed as functions of $\eps$ for $\eps^2 \in [10^{-5},0.04]$.  
As seen in Fig.~\ref{fig:U1U0rms}(a), we obtain $\trave{|U_1|} = \bigO(\eps^{\alpha'})$ for $\alpha' \approx 1.5$; that is, we find in agreement with \cite{Tana05} that the pattern mode $U_1$ indeed satisfies the MC scaling $\alpha'$=$\alpha$=3/2 \cite{MaCo00}.

\begin{figure}[tb]
  \begin{center}
  \includegraphics[width=3.2in]{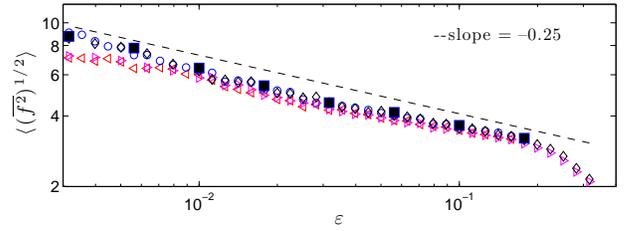} 
    \caption{(Color online) Scaling of averaged rms amplitudes of $\eps^{-2} U_0$ from \eqref{eq:nik} and $f$ from \eqref{eq:MCe1}--\eqref{eq:MCe2}.   This is the data of Fig.~\ref{fig:U1U0rms}(b) rescaled vertically by $\eps^{-2}$ to emphasize the anomalous scaling: according to the prediction of the Matthews-Cox scaling Ansatz, which for the large-scale mode is $f = \bigO(1)$, $U_0 = \bigO(\eps^2)$, these values are expected to be $\eps$-independent.\\
    Symbols are as in Fig.~\ref{fig:U1U0rms}.  The solid black squares are obtained from additional computations of the Nikolaevskiy PDE \eqref{eq:nik} for larger $L = \eps \ell = 2\pi \cdot 51.2$ ($m$=4).}
    \label{fig:frms}
  \end{center}
\end{figure}

The computed scaling of the reconstructed large-scale mode $U_0$ is rather more surprising: as shown in Fig.~\ref{fig:U1U0rms}(b), over a wide range $\eps^2 \in [10^{-5},0.04]$ our data is consistent with $\trave{U_0} = \bigO(\eps^{\beta'})$ with the anomalous exponent $\beta' \approx 1.75$, contrary to the prediction $\beta'$=$\beta$=2 from multiple-scale analysis.  
In the alternative representation of our data in Fig.~\ref{fig:frms} the deviation from the MC prediction is more readily apparent: $\eps^{-2} \trave{U_0}$ is clearly not $\eps$-independent, as confirmed also by some larger-domain computations performed as an additional check.

The leading-order amplitude equations \eqref{eq:MC1}--\eqref{eq:MC2}, being  $\eps$-independent, are manifestly unable to capture this unexpected scaling behavior.  
However, for comparison with the full Nikolaevskiy PDE, we have integrated the $\bigO(\eps)$ amplitude equations \eqref{eq:MCe1}--\eqref{eq:MCe2} for $A$ and $f$ as functions of the slow variables $X \in [0,L]$ and $T$, for the same $\eps$- and $L$-values as before.  The functions $u(x,t)$ reconstructed from $A$ and $f$ using \eqref{eq:genscal} are qualitatively similar to computed solutions of \eqref{eq:nik}; for $\eps > 0$ trajectories appear chaotic and are not observed to settle down.  

We extracted amplitude scaling behavior for \eqref{eq:MCe1}--\eqref{eq:MCe2} by computing the time-averaged $\rms$ amplitudes of $|\eps^{3/2} A|$ and $\eps^2 f$; as shown in Fig.~\ref{fig:U1U0rms}, the results agree remarkably well with those obtained for $|U_1|$ and $U_0$ from the full PDE \eqref{eq:nik}.  Equivalently, the statistics of the pattern mode $A$ are, unsurprisingly, asymptotically $\eps$-independent; but for the large-scale mode, we find that approximately $\eps^2 \trave{f} = \bigO(\eps^{7/4})$, or, as shown in Fig.~\ref{fig:frms}, $\trave{f} = \bigO(\eps^{-1/4})$: whereas one might have anticipated a small ($\bigO(\eps)$) effect of the added $\bigO(\eps)$ terms in the amplitude equations \eqref{eq:MCe1}--\eqref{eq:MCe2}, in fact they are able to capture the (asymptotically large) corrections to MC scaling.

We observe in Fig.~\ref{fig:frms} a possible weak dependence of $\trave{f}$ (and of $\eps^{-2} \trave{U_0}$) on the system size $L = \eps \ell$, which may merit further study; but that for each $L$ we have the same scaling with $\eps$ (with some possible finite-size effects appearing at the smaller domain size for $\eps^2 < 10^{-4}$).  The deviation for $\eps^2 > 0.04$, where the separation of scales is weak and the ``perturbation'' parameter is relatively large, is unsurprising.

The corrections to scaling appear not only in the rms amplitudes: 
While the multiple-scale Ansatz \eqref{eq:genscal} introduces only one slow time scale $T = \eps^2 t$, a glance at time series of Fourier coefficients $\hat{u}_k(t)$ computed from \eqref{eq:nik} for $k$$\approx$0 and $k$$\approx$1, or at the evolution of $U_0$ and $|U_1|$, suggests that for small $\eps$, at large scales the dynamics are much slower than for the pattern mode.

\begin{figure}[tb]
  \begin{center}
   \includegraphics[width=3.2in]{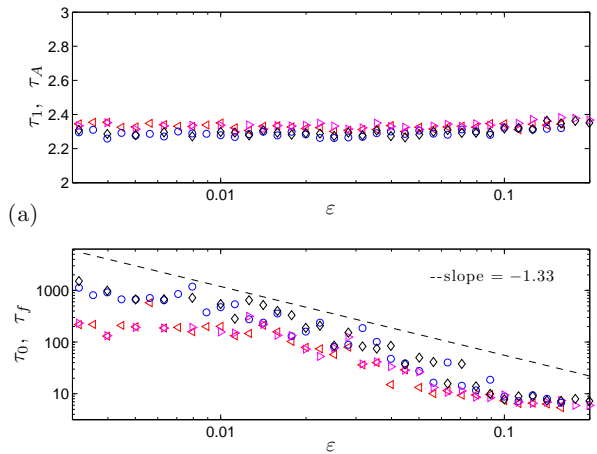} 
    \caption{(Color online) (a) Scaling of correlation times for the rms values of $|U_1|$ from \eqref{eq:nik}, and $|A|$ from \eqref{eq:MCe1}--\eqref{eq:MCe2}; we find $\tau_1 \approx \tau_A \approx 2.3$, decreasing slightly for larger $L$.  (b) Scaling of correlation times for the rms values of $U_0$ from \eqref{eq:nik} and $f$ from \eqref{eq:MCe1}--\eqref{eq:MCe2}.
    Symbols are as in Fig.~\ref{fig:U1U0rms}.}
    \label{fig:tau1tau0}
  \end{center}
\end{figure}

To begin to quantify this observation of potentially distinct time scales, for solutions of \eqref{eq:nik} we considered time series $\rsqave{U_0(\cdot,T)}$ and $\rsqave{|U_1(\cdot,T)|}$ of the $\rms$ mode amplitudes depending on the slow time variable $T$, and computed (full-width at half-maximum) autocorrelation times $\tau_0$, $\tau_1$ as functions of $\eps$.  Again for comparison, we similarly computed correlation times $\tau_f$, $\tau_A$ from time series of $\rms$ values of $f$, $|A|$ computed via \eqref{eq:MCe1}--\eqref{eq:MCe2}.  Our results are shown in Fig.~\ref{fig:tau1tau0}; again, the $\bigO(\eps)$ amplitude equations closely match the behavior of the full PDE. 

The asymptotic validity of the leading-order modulation equations \eqref{eq:MC1}--\eqref{eq:MC2} implies that these correlation times should be (asymptotically) $\eps$-independent.  Indeed, Fig.~\ref{fig:tau1tau0}(a) shows these predictions again satisfied for the pattern mode $U_1$: we find $\tau_1 \approx \tau_A \approx 2.3$, seemingly independent of $\eps$ (though with, again, an apparent weak dependence on domain size $L$).  However, as seen in Fig.~\ref{fig:tau1tau0}(b) the long-wave mode $U_0$ is substantially slower: while our statistics are not yet well-converged (particularly for small $\eps$), our data appear to indicate the presence of another time scale diverging as $\eps \to 0$, and scaling possibly as $\tau_0 \approx \tau_f \sim \eps^{-\delta}$, where our results suggest $\delta$ may be near 4/3.

\section{Discussion}
\label{sec:discussion}

The description that emerges for the Nikolaevskiy PDE \eqref{eq:nik} with sufficiently small $\eps$ is of spatiotemporally chaotic dynamics with strong scale separation and anomalous long-wave behavior.  The solution space features two distinct scaling regimes: The $\bigO(\eps)$ GL regime, which contains the roll solutions $u_q(x)$ together with their stable manifolds, is unstable to spatially varying perturbations, which lead to collapse to a smaller, apparently attracting, set in phase space, the $\bigO(\eps^{3/2})$ MC regime of ``Nikolaevskiy chaos''.  

Our observed anomalous scaling of the long-wave mode $U_0$ implies that the leading-order amplitude equations \eqref{eq:MC1}--\eqref{eq:MC2} do not fully describe the dynamics on the attractor of Nikolaevskiy STC, at least for the parameter range $\eps^2 \in [10^{-5},  0.04]$ we investigated.  This conclusion is supported by recent observations of non-extensive behavior in the Matthews-Cox equations \cite{SaTa07}.  

The higher-order modulation equations, however, do appear to capture the corrections to amplitude and time scaling: the function $f(X,T)$, evolving according to the coupled equations \eqref{eq:MCe1}--\eqref{eq:MCe2}, scales in the same way as $\eps^{-2} U_0$, where $U_0$ is the large-scale mode extracted from solutions of \eqref{eq:nik} by Fourier filtering.
Further investigation of \eqref{eq:MCe1}--\eqref{eq:MCe2} is needed to clarify the mechanism by which the addition of $\bigO(\eps)$ terms to the MC equations modifies the scaling.

\begin{acknowledgments}

This work was partially supported through an NSERC grant to R.W.  We would like to thank David Muraki for extensive valuable discussions and comments concerning this work.  R.W.\ would like to acknowledge helpful conversations with participants at the `Pattern Formation in Large Domains' program at the Newton Institute for Mathematical Sciences, Cambridge.
We also thank Youngsuk Lee for useful discussions,
and the IRMACS Centre at Simon Fraser University for its support.

\end{acknowledgments}

\newcommand{\SortNoop}[1]{}


\begin{thebibliography}{26}
\expandafter\ifx\csname natexlab\endcsname\relax\def\natexlab#1{#1}\fi
\expandafter\ifx\csname bibnamefont\endcsname\relax
  \def\bibnamefont#1{#1}\fi
\expandafter\ifx\csname bibfnamefont\endcsname\relax
  \def\bibfnamefont#1{#1}\fi
\expandafter\ifx\csname citenamefont\endcsname\relax
  \def\citenamefont#1{#1}\fi
\expandafter\ifx\csname url\endcsname\relax
  \def\url#1{\texttt{#1}}\fi
\expandafter\ifx\csname urlprefix\endcsname\relax\def\urlprefix{URL }\fi
\providecommand{\bibinfo}[2]{#2}
\providecommand{\eprint}[2][]{\url{#2}}

\bibitem[{\citenamefont{Cross and Hohenberg}(1993)}]{CrHo93}
\bibinfo{author}{\bibfnamefont{M.}~\bibnamefont{Cross}} \bibnamefont{and}
  \bibinfo{author}{\bibfnamefont{P.}~\bibnamefont{Hohenberg}},
  \bibinfo{journal}{Rev. Mod. Phys.} \textbf{\bibinfo{volume}{65}},
  \bibinfo{pages}{851} (\bibinfo{year}{1993}).

\bibitem[{\citenamefont{Hoyle}(2006)}]{Hoyl06k}
\bibinfo{author}{\bibfnamefont{R.}~\bibnamefont{Hoyle}},
  \emph{\bibinfo{title}{Pattern Formation: An introduction to methods}}
  (\bibinfo{publisher}{Cambridge University Press},
  \bibinfo{address}{Cambridge}, \bibinfo{year}{2006}).

\bibitem[{\citenamefont{Rabinovich et~al.}(2000)\citenamefont{Rabinovich,
  Ezersky, and Weidman}}]{REW00k}
\bibinfo{author}{\bibfnamefont{M.~I.} \bibnamefont{Rabinovich}},
  \bibinfo{author}{\bibfnamefont{A.~B.} \bibnamefont{Ezersky}},
  \bibnamefont{and} \bibinfo{author}{\bibfnamefont{P.~D.}
  \bibnamefont{Weidman}}, \emph{\bibinfo{title}{The Dynamics of Patterns}}
  (\bibinfo{publisher}{World Scientific}, \bibinfo{address}{Singapore},
  \bibinfo{year}{2000}).

\bibitem[{\citenamefont{Fishman and Egolf}(2006)}]{FiEg06}
\bibinfo{author}{\bibfnamefont{M.~P.} \bibnamefont{Fishman}} \bibnamefont{and}
  \bibinfo{author}{\bibfnamefont{D.~A.} \bibnamefont{Egolf}},
  \bibinfo{journal}{Phys. Rev. Lett.} \textbf{\bibinfo{volume}{96}},
  \bibinfo{pages}{054103} (\bibinfo{year}{2006}).

\bibitem[{\citenamefont{Bernoff}(1994)}]{Bern94}
\bibinfo{author}{\bibfnamefont{A.~J.} \bibnamefont{Bernoff}},
  \bibinfo{journal}{Euro. J. Appl. Math.} \textbf{\bibinfo{volume}{5}},
  \bibinfo{pages}{267} (\bibinfo{year}{1994}).

\bibitem[{\citenamefont{Bernoff et~al.}(1995)\citenamefont{Bernoff, Kuske,
  Matkowsky, and Volpert}}]{BKMV95}
\bibinfo{author}{\bibfnamefont{A.~J.} \bibnamefont{Bernoff}},
  \bibinfo{author}{\bibfnamefont{R.}~\bibnamefont{Kuske}},
  \bibinfo{author}{\bibfnamefont{B.~J.} \bibnamefont{Matkowsky}},
  \bibnamefont{and} \bibinfo{author}{\bibfnamefont{V.}~\bibnamefont{Volpert}},
  \bibinfo{journal}{SIAM J. Appl. Math.} \textbf{\bibinfo{volume}{55}},
  \bibinfo{pages}{485} (\bibinfo{year}{1995}).

\bibitem[{\citenamefont{Tribel'ski\u{\i}}(1997)}]{Trib97}
\bibinfo{author}{\bibfnamefont{M.~I.} \bibnamefont{Tribel'ski\u{\i}}},
  \bibinfo{journal}{Usp. Fiz. Nauk} \textbf{\bibinfo{volume}{167}},
  \bibinfo{pages}{167} (\bibinfo{year}{1997}).

\bibitem[{\citenamefont{Nikolaevskii}(1989)}]{Niko89}
\bibinfo{author}{\bibfnamefont{V.~N.} \bibnamefont{Nikolaevskii}}, in
  \emph{\bibinfo{booktitle}{Recent Advances in Engineering Science}}
  (\bibinfo{publisher}{Springer-Verlag}, \bibinfo{address}{Berlin},
  \bibinfo{year}{1989}), vol.~\bibinfo{volume}{39} of
  \emph{\bibinfo{series}{Lecture Notes in Engineering}}, pp.
  \bibinfo{pages}{210--221}.

\bibitem[{\citenamefont{Beresnev and Nikolaevskiy}(1993)}]{BeNi93}
\bibinfo{author}{\bibfnamefont{I.~A.} \bibnamefont{Beresnev}} \bibnamefont{and}
  \bibinfo{author}{\bibfnamefont{V.~N.} \bibnamefont{Nikolaevskiy}},
  \bibinfo{journal}{Physica D} \textbf{\bibinfo{volume}{66}},
  \bibinfo{pages}{1} (\bibinfo{year}{1993}).

\bibitem[{\citenamefont{Fujisaka and Yamada}(2001)}]{FuYa01}
\bibinfo{author}{\bibfnamefont{H.}~\bibnamefont{Fujisaka}} \bibnamefont{and}
  \bibinfo{author}{\bibfnamefont{T.}~\bibnamefont{Yamada}},
  \bibinfo{journal}{Prog. Theor. Phys.} \textbf{\bibinfo{volume}{106}},
  \bibinfo{pages}{315} (\bibinfo{year}{2001}).

\bibitem[{\citenamefont{Tanaka}(2004)}]{Tana04}
\bibinfo{author}{\bibfnamefont{D.}~\bibnamefont{Tanaka}},
  \bibinfo{journal}{Phys. Rev. E} \textbf{\bibinfo{volume}{70}},
  \bibinfo{pages}{015202(R)} (\bibinfo{year}{2004}).

\bibitem[{\citenamefont{Cox and Matthews}(2007)}]{CoMa07}
\bibinfo{author}{\bibfnamefont{S.~M.} \bibnamefont{Cox}} \bibnamefont{and}
  \bibinfo{author}{\bibfnamefont{P.~C.} \bibnamefont{Matthews}},
  \bibinfo{journal}{Phys. Rev. E} \textbf{\bibinfo{volume}{76}},
  \bibinfo{pages}{056202} (\bibinfo{year}{2007}).

\bibitem[{\citenamefont{Matthews and Cox}(2000)}]{MaCo00}
\bibinfo{author}{\bibfnamefont{P.~C.} \bibnamefont{Matthews}} \bibnamefont{and}
  \bibinfo{author}{\bibfnamefont{S.~M.} \bibnamefont{Cox}},
  \bibinfo{journal}{Phys. Rev. E} \textbf{\bibinfo{volume}{62}},
  \bibinfo{pages}{R1473} (\bibinfo{year}{2000}).

\bibitem[{\citenamefont{Tribelsky and Velarde}(1996)}]{TrVe96}
\bibinfo{author}{\bibfnamefont{M.~I.} \bibnamefont{Tribelsky}}
  \bibnamefont{and} \bibinfo{author}{\bibfnamefont{M.~G.}
  \bibnamefont{Velarde}}, \bibinfo{journal}{Phys. Rev. E}
  \textbf{\bibinfo{volume}{54}}, \bibinfo{pages}{4973} (\bibinfo{year}{1996}).

\bibitem[{\citenamefont{Tribelsky}(2008)}]{Trib08}
\bibinfo{author}{\bibfnamefont{M.~I.} \bibnamefont{Tribelsky}},
  \bibinfo{journal}{Phys. Rev. E} \textbf{\bibinfo{volume}{77}},
  \bibinfo{pages}{035202(R)} (\bibinfo{year}{2008}).

\bibitem[{\citenamefont{Tribelsky and Tsuboi}(1996)}]{TrTs96}
\bibinfo{author}{\bibfnamefont{M.~I.} \bibnamefont{Tribelsky}}
  \bibnamefont{and} \bibinfo{author}{\bibfnamefont{K.}~\bibnamefont{Tsuboi}},
  \bibinfo{journal}{Phys. Rev. Lett.} \textbf{\bibinfo{volume}{76}},
  \bibinfo{pages}{1631} (\bibinfo{year}{1996}).

\bibitem[{\citenamefont{Rossberg et~al.}(1996)\citenamefont{Rossberg, Hertrich,
  Kramer, and Pesch}}]{RHKP96}
\bibinfo{author}{\bibfnamefont{A.~G.} \bibnamefont{Rossberg}},
  \bibinfo{author}{\bibfnamefont{A.}~\bibnamefont{Hertrich}},
  \bibinfo{author}{\bibfnamefont{L.}~\bibnamefont{Kramer}}, \bibnamefont{and}
  \bibinfo{author}{\bibfnamefont{W.}~\bibnamefont{Pesch}},
  \bibinfo{journal}{Phys. Rev. Lett.} \textbf{\bibinfo{volume}{76}},
  \bibinfo{pages}{4729} (\bibinfo{year}{1996}).

\bibitem[{\citenamefont{Hidaka et~al.}(1997)\citenamefont{Hidaka, Huh, Hayashi,
  Kai, and Tribelsky}}]{HHHKT97}
\bibinfo{author}{\bibfnamefont{Y.}~\bibnamefont{Hidaka}},
  \bibinfo{author}{\bibfnamefont{J.-H.} \bibnamefont{Huh}},
  \bibinfo{author}{\bibfnamefont{K.-i.} \bibnamefont{Hayashi}},
  \bibinfo{author}{\bibfnamefont{S.}~\bibnamefont{Kai}}, \bibnamefont{and}
  \bibinfo{author}{\bibfnamefont{M.~I.} \bibnamefont{Tribelsky}},
  \bibinfo{journal}{Phys. Rev. E} \textbf{\bibinfo{volume}{56}},
  \bibinfo{pages}{R6256} (\bibinfo{year}{1997}).

\bibitem[{\citenamefont{Cox and Matthews}(2002)}]{CoMa02}
\bibinfo{author}{\bibfnamefont{S.~M.} \bibnamefont{Cox}} \bibnamefont{and}
  \bibinfo{author}{\bibfnamefont{P.~C.} \bibnamefont{Matthews}},
  \bibinfo{journal}{J. Comput. Phys.} \textbf{\bibinfo{volume}{176}},
  \bibinfo{pages}{430} (\bibinfo{year}{2002}).

\bibitem[{\citenamefont{Kassam and Trefethen}(2005)}]{KaTr03}
\bibinfo{author}{\bibfnamefont{A.-K.} \bibnamefont{Kassam}} \bibnamefont{and}
  \bibinfo{author}{\bibfnamefont{L.~N.} \bibnamefont{Trefethen}},
  \bibinfo{journal}{SIAM J. Sci. Comput.} \textbf{\bibinfo{volume}{26}},
  \bibinfo{pages}{1214} (\bibinfo{year}{2005}).

\bibitem[{\citenamefont{Tanaka}(2005)}]{Tana05}
\bibinfo{author}{\bibfnamefont{D.}~\bibnamefont{Tanaka}},
  \bibinfo{journal}{Phys. Rev. E} \textbf{\bibinfo{volume}{71}},
  \bibinfo{pages}{025203(R)} (\bibinfo{year}{2005}).

\bibitem[{\citenamefont{Wittenberg and Holmes}(1999)}]{WiHo99}
\bibinfo{author}{\bibfnamefont{R.~W.} \bibnamefont{Wittenberg}}
  \bibnamefont{and} \bibinfo{author}{\bibfnamefont{P.}~\bibnamefont{Holmes}},
  \bibinfo{journal}{Chaos} \textbf{\bibinfo{volume}{9}}, \bibinfo{pages}{452}
  (\bibinfo{year}{1999}).

\bibitem[{\citenamefont{Coullet and Fauve}(1985)}]{CoFa85}
\bibinfo{author}{\bibfnamefont{P.}~\bibnamefont{Coullet}} \bibnamefont{and}
  \bibinfo{author}{\bibfnamefont{S.}~\bibnamefont{Fauve}},
  \bibinfo{journal}{Phys. Rev. Lett.} \textbf{\bibinfo{volume}{55}},
  \bibinfo{pages}{2857} (\bibinfo{year}{1985}).

\bibitem[{\citenamefont{Malomed}(1992)}]{Malo92}
\bibinfo{author}{\bibfnamefont{B.~A.} \bibnamefont{Malomed}},
  \bibinfo{journal}{Phys. Rev. A} \textbf{\bibinfo{volume}{45}},
  \bibinfo{pages}{1009} (\bibinfo{year}{1992}).

\bibitem[{\citenamefont{Fujisaka et~al.}(2003)\citenamefont{Fujisaka, Honkawa,
  and Yamada}}]{FHY03}
\bibinfo{author}{\bibfnamefont{H.}~\bibnamefont{Fujisaka}},
  \bibinfo{author}{\bibfnamefont{T.}~\bibnamefont{Honkawa}}, \bibnamefont{and}
  \bibinfo{author}{\bibfnamefont{T.}~\bibnamefont{Yamada}},
  \bibinfo{journal}{Prog. Theor. Phys.} \textbf{\bibinfo{volume}{109}},
  \bibinfo{pages}{911} (\bibinfo{year}{2003}).

\bibitem[{\citenamefont{Sakaguchi and Tanaka}(2007)}]{SaTa07}
\bibinfo{author}{\bibfnamefont{H.}~\bibnamefont{Sakaguchi}} \bibnamefont{and}
  \bibinfo{author}{\bibfnamefont{D.}~\bibnamefont{Tanaka}},
  \bibinfo{journal}{Phys. Rev. E} \textbf{\bibinfo{volume}{76}},
  \bibinfo{pages}{025201(R)} (\bibinfo{year}{2007}).

\end{thebibliography}
\end{document}